\documentstyle[12pt]{article}
 \def\be{\begin{equation}}
 \def\ee{\end{equation}}
 \def\bea{\begin{eqnarray}}
 \def\eea{\end{eqnarray}}
 \def\nn{\nonumber}
 \def\vx{{\vec x}}
 \def\vk{{\vec k}}
 \def\vy{{\vec y}}
 \def\vz{{\vec z}}
 \def\p{\phi}

\begin{document}
 \title{dS/CFT Correspondence from a Holographic Description of Massless Scalar Fields in
 Minkowski Space-Time}
 \author{Farhang Loran\thanks{e-mail:
loran@cc.iut.ac.ir}\\ \\
  {\it Department of  Physics, Isfahan University of Technology (IUT)}\\
{\it Isfahan,  Iran,} }
\date{}
\maketitle

 \begin{abstract} We solve Klein-Gordon equation for massless scalars on d+1 dimensional
 Minkowski (Euclidean) space in terms of the Cauchy data on the hypersurface
 t=0. By inserting the solution into the action of massless scalars in Minkowski
 (Euclidean) space we obtain the action of dual theory on the boundary
 t=0 which is exactly the holographic dual of conformally coupled
 scalars on d+1 dimensional (Euclidean anti) de Sitter space obtained in (A)dS/CFT
 correspondence. The observed equivalence of dual theories is explained using
 the one-to-one map between conformally coupled scalar fields on Minkowski (Euclidean) space and (Euclidean anti) de
 Sitter space which is an isomorphism between the hypersurface t=0 of
 Minkowski (Euclidean) space and the boundary of (A)dS space.
 \end{abstract}

\section{Introduction}
 The holographic principle is in general a duality between the physics of a generic
 space-time and a theory on the boundary (the holographic screen) \cite{Hol}.
 Quantitative examples of the holographic principle are AdS/CFT \cite{Hen,Witten} and
 dS/CFT \cite{Strom,Park} correspondence. Recently de Boer and Solodukhin proposed a
 holographic description of Minkowski space in terms of a dual CFT defined on the boundary
 of the light cone \cite{Sol}.
 \par
 In this paper we give a holographic description of massless scalar
 field on d+1 dimensional Minkowski (Euclidean) space in terms of a dual theory
 on the  hypersurface $t=0$. The dual action is obtained by inserting the solution of
 Klein-Gordon equation into the action of massless
 scalars on Minkowski (Euclidean) space. As is shown, the action on the boundary is
 equivalent to the action of the theory on the boundary of (Euclidean anti) de Sitter space
 dual to conformally coupled scalars obtained in (A)dS/CFT correspondence.
 The equivalence of dual theories can be  explained by using the injective map between
 massless scalars on Minkowski (Euclidean) space and dS (Euclidean AdS) space with
 the same dimensionality which induces an isomorphism
 between the hypersurface t=0 of Minkowski (Euclidean) space and the boundary of
 (Euclidean anti) de Sitter space.
 \par
 The organization of the paper is as follows. In section 2 we study the correspondence
 between massless scalars in $d+1$ dimensional Euclidean space $R^{d+1}$ and
 scalars with mass $m^2=\frac{1-d^2}{4}$ (the conformally coupled scalars) in Euclidean
 $\mbox{AdS}_{d+1}$. Section 3 is devoted to scalars on Minkowski space and
 dS/CFT correspondence. The notation used in sections 2 and 3 are similar to
 the  notation of references \cite{Witten} and \cite{Strom}
 respectively. In section 3, we describe the ${\cal O}^-$ region
 of $\mbox{dS}_{d+1}$ by metric, $ds^2=t^{-2}(-dt^2+dx_i^2)$ which is
 related to the metric $ds^2=-du^2+e^{-2u}dx_i^2 $ used in \cite{Strom} by the identity
 $t=e^u$. Most significant properties of dS space related to dS/CFT correspondence
 are studied in \cite{Les}. Various aspects of AdS/CFT and dS/CFT duality are recently
 reviewed in \cite{Boer}. In section 4, we summarize our results
 and address some of applications of the introduced method. The scalar field theory in
 curved spacetime is briefly reviewed in the appendix.
\section{Massless Scalars on Euclidean Space}
 The equation of motion for massive scalar fields on Euclidean
 $\mbox{AdS}_{d+1}$ with metric,
 \be
  ds^2=\frac{1}{t^2}\left(dt^2+\sum_{i=1}^d dx_i^2\right),
 \label{met}
 \ee
 is
 \be
 \left(t^2\partial_t^2+(1-d)t\partial_t+t^2\nabla^2-m^2\right)\Phi=0,
 \label{a1}
 \ee
 where $\partial_t=\frac{\partial}{\partial t}$ and
 \be
 \nabla^2=\sum_{i=1}^d\frac{\partial^2}{\partial x_i^2}.
 \ee
 See the appendix for a brief review of scalar field theory in curved spacetimes.
 One can easily show that if $\phi$ is a massless scalar
 field in $d+1$ dimensional Euclidean space $R^{d+1}$ with metric
 $ds^2=dt^2+dx_i^2,$
 satisfying the equation
 \be
 (\partial_t^2+\nabla^2)\phi=0,
 \label{a2}
 \ee
 then
 \be
 \Phi=t^{\frac{d-1}{2}}\phi,
 \label{map}
 \ee
 is a solution of Eq.(\ref{a1}) with mass
 $m^2=\frac{1-d^2}{4}$. Since $\frac{-d^2}{4}<m^2<0$, this solution is
 stable in $AdS_{d+1}$.
 From AdS/CFT correspondence \cite{Witten} it is known that the dual theory on the
 boundary $t=0$ is a conformal theory with the following action,
 \be
 I[\phi]=\int d^dy d^dz
 \frac{\phi_0({\vec y})\phi_0({\vec z})}{\left|{\vec y}-{\vec
 z}\right|^{d+\lambda_+}},
 \label{a3}
 \ee
 where $\lambda_+$
 is the larger root of the equation $\lambda(\lambda+d)=m^2$. Here
 $\lambda_+=\frac{(1-d)}{2}$ as far as $m^2=\frac{(1-d^2)}{4}$.
 $\phi_0$ is a function on the boundary such that
 $\Phi(\vx,t)\sim t^{-\lambda_+}\phi_0$ as $t\to 0$. From the map (\ref{map}),
 one can interpret $\phi_0(\vx)$ in
 Euclidean space  $R^{d+1}$ as the initial data on the hypersurface $t=0$.
 Therefore one expects that the action (\ref{a3})
 can be obtained from the action of scalar fields in $d+1$ dimensional
 Euclidean space,
 \be
 I[\phi]=\frac{1}{2}\int
 dtd^dx\left((\partial_t\phi)^2+(\nabla\phi)^2\right),
 \label{EA}
 \ee
 if one solves equation (\ref{a2})
 in terms of the initial data $\phi_0(\vx)$ given on the
 hypersurface $t=0$. Proof is as follows:
 \par
 The most general solution of the equation of motion that vanishes as $t$
 tends to infinity is
 \be
 \phi(\vx,t)=\int d^dk {\tilde \phi}(\vk)e^{i\vk.\vx}e^{-\omega
 t},
 \label{a4}
 \ee
 where $\omega=\left|\vk\right|$ and
 \be
 {\tilde \phi}(\vk)=\int d^dx \phi_0(\vx)e^{-i\vk.\vx}.
 \label{a41}
 \ee
 Inserting (\ref{a41}) into (\ref{a4}) one obtains
 \be
 \p(\vx,t)=\int d^dy\ {\cal G}(\vx,t;\vy)\p_0(\vy),
 \label{a42}
 \ee
 where
 \be
 {\cal G}(\vx,t;\vy)=\int d^dk\  e^{-\omega t} e^{i\vk.(\vx-\vy)}.
 \label{g}
 \ee
 ${\cal G}(\vx,t;\vy)$ is the solution of wave equation i.e. $\Box {\cal G}=0$,
 with the initial condition ${\cal G}(\vx,0;\vy)=\delta^d(\vx-\vy)$.
 To obtain the action of the corresponding theory on the boundary $t=0$ one should insert
 (\ref{a42}) into (\ref{EA}). But it is more suitable to rewrite the action (\ref{EA}) in
 the form,
 \be
 I[\phi]=-\frac{1}{2}\int d^{d+1}x\ \p\Box\p-\frac{1}{2}\int d^dx\
 \p_0(\vx)\partial_t\p_0(\vx),
 \label{intbypart}
 \ee
 which is obtained by an integration by part and under the assumption that
 $\phi(x)$ vanishes as $t$ tends to infinity and also at spatial
 infinity. Inserting (\ref{a42}) into (\ref{intbypart}), the first term
 vanishes and from the second term one obtains:
 \be
 I[\phi]=\frac{1}{2}\int d^dyd^dz\ \phi_0({\vec y})\phi_0({\vec z})F({\vec y}-{\vec
 z}),
 \label{a5}
 \ee
 in which
 \be
 F(\vx)=\int d^dk\  \omega e^{i\vk.\vx}.
 \ee
 As can be verified from the rotational invariance ($\vx\to {\bf R}\vx$,
 ${\bf R}\in SO(d)$),
 $F(\vx)$ depends only on the norm of $\vx$. By scaling $\vx$ by a factor $\lambda>0$ one
 can also show that $F(\lambda\vx)=\lambda^{-(d+1)}F(\vx)$ and consequently,
 \be
 \vx.\nabla F(\vx)=\lim_{\lambda\to 1}\frac{F(\lambda\left|\vx\right|)-F(\left|\vx\right|)}
 {\lambda-1}=-(d+1)F(\left|\vx\right|).
 \ee
 Therefore, $F(\vx)=\mbox{Const.}\left|\vx\right|^{-(d+1)}$ and
 \be
 I[\phi]=\mbox{Const.}\int d^dyd^dz\ \frac{\phi_0({\vec y})\phi_0({\vec
 z})}{ \left|{\vec y}-{\vec z}\right|^{d+1}},
 \label{af}
 \ee
 which is equal to (\ref{a3}) derived in AdS/CFT correspondence.
 \par
  One should note that in addition to the general solution
 (\ref{a4}), Eq.(\ref{a2}) has one further solution $\phi(\vx,t)=\alpha
 t$ where $\alpha$ is some constant that can not be determined from the initial data
 $\phi_0(\vx)$. Although this solution is meaningless in Euclidean
 space, but $\alpha$ is equal to the value of $\Phi(\vx,t)$ at $t\to \infty$ which is an
 additional point on the boundary of AdS.
 \par
 One can interpret the above result as a holographic
 principle for Euclidean space. But our analysis considers only
 massless scalars. As can be easily verified, it is not possible to generalize the map
 (\ref{map}) to include massive scalar fields in $R^{d+1}$. A
 reason for this is the fact that the equation of motion of scalar
 fields $(\Box +m^2)\phi=0$ is not conformally covariant unless
 $m=0$.
 \section{Massless Scalars on Minkowski Space-time}
 Similar to section 2, one can show that massless scalar fields in
 $d+1$ dimensional Minkowski space-time $M_{d+1}$ can be mapped
 by (\ref{map}) to scalars with mass $m^2=\frac{d^2-1}{4}$
 on $dS_{d+1}$. To verify this claim one can use the following
 metrics for $dS_{d+1}$ and $M_{d+1}$ respectively:
 \be
 ds^2_{dS}=\frac{1}{t^2}\left(-dt^2+\sum_{i=1}^d
 dx_i^2\right),
 \label{b1}
 \ee
 \be
 ds^2_{M}=\left(-dt^2+\sum_{i=1}^d
 dx_i^2\right)
 \label{b2}
 \ee
 The metric (\ref{b1}) covers only half of dS space. This region called ${\cal O}^-$ is the
 region observed by an observer on the south pole ${\cal I}^-$
 but is behind the horizon of the
 observer on the north pole ${\cal I}^+$. By construction  $t>0$.
 Following the Strominger proposal, dual operators living on the
 boundary ${\cal I}^-$ obey,
 \be
 \left<{\cal O}_\phi(z,{\bar z}),{\cal O}_\phi(v,{\bar
 v})\right>=\frac{\mbox{const.}}{\left|z-v\right|^{2h_+}},
 \label{b3}
 \ee
 where
 \be
 h_\pm=\frac{1}{2}\left(d\pm\sqrt{d^2-4m^2}\right).
 \label{b4}
 \ee
 Again, the existence of the map (\ref{map}) suggests that
 Eq.(\ref{b3}) can be obtained by solving the equations of motion of massless scalar
 fields on $M_{d+1}$ in terms of the initial data at $t=0$. If yes
 then the final result can be interpreted  as a holographic
 description of massless scalars in Minkowski space time. Such a
 description can be made covariant  by considering a covariant boundary \cite{Busso}
 instead of the hypersurface $t=0$, which here corresponds to the ${\cal I}^-$ by
 (\ref{map}).
 \par
 General arguments \cite{Strom} show that a massive scalar field
 behaves as $t^{h_\pm}\phi_\pm$ near ${\cal I}^-$ and the dual theory on the boundary
 ${\cal I}^-$ (the planar past region) is described by the action
 \be
 I[\phi]=\int_{{\cal I}^-}d^dyd^d(z)\left(\frac{\phi_-({\vec
 y})\phi_-({\vec z})}{\left|{\vec y}-{\vec z}\right|^{2h_+}}+
 \frac{\phi_+({\vec
 y})\phi_+({\vec z})}{\left|{\vec y}-{\vec
 z}\right|^{2h_-}}\right).
 \label{int1}
 \ee
 Since $h_-=\frac{d-1}{2}$ for
 $m^2=\frac{d^2-1}{4}$, using Eq.(\ref{map}) one verifies that,
 $\phi_-(\vx)=\phi(\vx,t)|_{t=0}$. As will be exactly
 shown, $\phi_+(\vx)=i\partial_t\phi(\vx,t)|_{t=0}$.
 Two evidences for this claim are:
 \begin{enumerate}
 \item{Since $h_+=h_-+1$ for $m^2=\frac{d^2-1}{4}$, using Eq.(\ref{map})
 one can verify that $\partial_t\phi$ mapped to $dS_{d+1}$
 behaves as $t^{h_+}$ near ${\cal I}^-$ as demanded.}
 \item{A general solution of equation of motion for massless scalars in $M_{d+1}$,
 contains oscillating terms with both positive and
 negative frequencies. Therefore $\phi(\vx,t)$ can only be given in terms of both
 $\phi_0(\vx,0)$ and $\partial_t\phi_0(\vx,0)$.}
 \end{enumerate}
 \par
 In other words the main purpose of this section is to derive the
 action (\ref{int1}) of the dual theory by inserting the solution of Klein-Gordon
 equation for massless scalar fields in $M_{d+1}$,
 \be
 \left(\partial_t^2-\nabla^2\right)\phi=0,
 \label{KG}
 \ee
 satisfying the initial conditions
 \be
 \phi(\vx,0)=\phi_-(\vx),
 \hspace{1cm}\partial_t\phi(\vx,0)=i\phi_+(\vx),
 \label{IC}
 \ee
 into the action of massless scalar fields in $M_{d+1}$,
 \be
 I[\phi]=\frac{1}{2}\int
 dtd^dx\left((\partial_t\phi)^2-(\nabla\phi)^2\right).
 \label{MA}
 \ee
 The most general solution of Eq.(\ref{KG}) satisfying the initial conditions (\ref{IC}) is
 \bea
 \phi(\vx,t)&=&\frac{1}{2}\int d^dyd^dk
 \left(\phi_-(\vy)-\frac{\phi_+(\vy)}{\omega}\right)
 e^{i\vk.(\vx-\vy)}e^{-i\omega t}\nn\\
 &+& \frac{1}{2}\int d^dyd^dk \left(\phi_-(\vy)+\frac{\phi_+(\vy)}{\omega}\right)
 e^{i\vk.(\vx-\vy)}e^{i\omega t},
 \eea
 where $\omega=\left|\vk\right|$. Inserting this solution into
 (\ref{MA}),  one obtains that up to some constant coefficient,
 \be
 I[\phi]=
 \frac{1}{2}\int d^dyd^dz {\em Re}[G(\vy,\vz)]
 \label{b6}
 \ee
 where
 \be
 G(\vy,\vz)=-\int d^dk \left(\omega\phi_-(\vy)+\phi_+(\vy)\right)
 \left(\omega\phi_-(\vz)+\phi_+(\vz)\right)
 e^{i\vk.(\vz-\vy)}f(\omega)
 \ee
 in which,
 \bea
 f(\omega)&=&\lim_{\alpha\to ^+0}\int_0^\infty dt
 e^{-(\alpha+2i\omega)t}\nn\\
 &=&\lim_{\alpha\to ^+0}\frac{1}{\alpha+2i\omega}
 \eea
 To obtain ${\em Re}[G(\vy,\vz)]$, one should note that
 $f(\omega)=f^*(\omega)$. To prove this equality one can show that
 for any analytic function $g(\omega)$
 \bea
 \int_{-\infty}^\infty d\omega g(\omega)f(\omega)&=&\int_{-\infty}^\infty d\omega
 g(\omega)f^*(\omega)\nn\\
 &=&\lim_{\alpha\to ^+0}(-2\pi) g(\alpha).
 \eea
 Consequently, up to some constant coefficients,
 \bea
 {\em Re}[G(\vy,\vz)]&=&\int d^dk
 \frac{\omega^2\phi_0(\vy)\phi_-(\vz)+\phi_+(\vy)\phi_+(\vz)}{\omega}e^{i\vk.(\vz-\vy)}\nn\\
 &=&\mbox{const.}\left(\frac{\phi_-(\vy)\phi_-(\vz)}{\left|\vz-\vy\right|^{d+1}}+
 \frac{\phi_+(\vy)\phi_+(\vz)}{\left|\vz-\vy\right|^{d-1}}\right).
 \eea
 \section{Conclusion}
 Considering the hypersurface t=0 as the boundary (the holographic screen)
 of  Minkowski (Euclidean) space, we obtained the action of dual
 theory on the boundary by solving the Klein-Gordon equation of
 motion in terms of the Cauchy data and inserting the solution into
 the action of massless scalar fields in the bulk.
 Since massless scalars on Minkowski (Euclidean)
 space are in one-to-one correspondence to conformally coupled scalars on (Euclidean anti)
 de Sitter space with the same dimension one expects that (as is verified) the dual
 action is equal to the dual theory on the boundary of (A)dS
 space obtained in (A)dS/CFT correspondence.
 \par
 Using this method one can study (A)dS/CFT duality in the
 case of self interacting conformally coupled scalar field theories, i.e D=3 $\p^6$-model,
 D=4 $\p^4$-model and D=6 $\p^3$, by solving the corresponding non-linear wave equation
 (by perturbation) and obtaining the action of the dual theory on the boundary by
 inserting the solution into the action of massless scalars on Minkowski (Euclidean) space
 (the bulk). Furthermore the same method can be used to study (A)dS/CFT
 correspondence and the holographic description of Minkowski
 space-time in the case of higher spin fields \cite{Spinor}.
 \section{Appendix}
 In this appendix we briefly review scalar field theory in $D=d+1$ dimensional
 curved spacetime.  The action for the scalar field $\phi$ is
 \be
 S=\int d^Dx\;
 \sqrt{\left|g\right|}\frac{1}{2}\left(g^{\mu\nu}\partial_\mu\phi\partial_\nu\phi -(m^2+\xi
 R)\phi^2\right),
 \ee
 for which the equation of motion is
 \be
 \left(\Box + m^2+\xi R\right)\phi=0,\hspace{1cm}
 \Box=|g|^{-1/2}\partial_\mu\left|g\right|^{1/2}g^{\mu\nu}\partial_\nu.
 \ee
 (With $\hbar$ explicit, the mass $m$ should be replaced by
 $m/\hbar$.) The case with $m=0$ and $\xi=\frac{d-1}{4d}$ is referred to
 as conformal coupling  \cite{Ted}.
 \par
 Using Eq.(\ref{b1}) it is easy to show that the Ricci scalar $R$ for $\mbox{dS}_{d+1}$
 space is $R=d(d+1)$ where we have set the dS radius $\ell=1$. Therefore, the action for
 conformally coupled scalars in $\mbox{dS}_{d+1}$ is
 \be
 S=\int d^Dx\;
 \sqrt{\left|g\right|}\frac{1}{2}\left(g^{\mu\nu}\partial_\mu\phi\partial_\nu\phi -
 (\frac{d^2-1}{4})\phi^2\right),
 \ee
 Similar result can be obtained for AdS space using Eq.(\ref{met}).

\end{document}